\documentclass[aps,twocolumn,amsmath,amsfonts,amssymb]{revtex4}
\usepackage{graphicx}
\usepackage{epsf}
\usepackage{epstopdf}

\begin{document}

\title{Matter-wave Fractional Revivals in a Ring Waveguide}

\author{Jayanta Bera$^{\mathrm{1}}$}
\author{Suranjana Ghosh$^{\mathrm{2}}$}
\author{Luca Salasnich$^{\mathrm{3}}$}
\author{Utpal Roy$^{\mathrm{1}}$ \footnote{uroy@iitp.ac.in}}

\affiliation{$^{\mathrm{1}}$Indian Institute of Technology Patna, Bihta, Patna-801106, India \\$^{\mathrm{2}}$Indian Institute of Science Education and Research Kolkata, West Bengal, India \\$^{\mathrm{3}}$ Dipartimento di Fisica e Astronomia ``Galileo Galilei", Universit\'{a} di Padova, Via Marzolo 8, 35131 Padova, Italy}

\begin{abstract}
We report fractional revivals phenomena in ultracold matter-wave inside a ring-waveguide. The specific fractional revival times are precisely identified and corresponding spatial density patterns are depicted. Thorough analyses of the autocorrelation function and quantum carpet provide clear evidences of their occurrence. The exhibited theoretical model is in exact conformity of our numerical results. We also investigate the stability of the condensate and a variation of revival time with the diameter of the ring.
\end{abstract}

\maketitle

\section{Introduction}
Revival and fractional revivals (FR) are very rich quantum phenomena which appear during the long time evolution of a well localized wave-packet \cite{R1,R2}. For some typical systems, wave-packets initially disperse in the course of its time evolution and subsequently reunite quite systematically at some specific evolution times, which are called FR instances. FR mostly appear in quantum systems with nonlinear energy spectrum: Rydberg wave packet \cite{R3}, infinite square well \cite{R4}, molecular systems \cite{R5,R6,R7}, photon cavity systems \cite{R8}, chaotic light \cite{R9}, just to mention a few. FR have found profound applications towards quantum information and quantum computation: quantum logic gates \cite{R10}, Renyi uncertainty relations \cite{R11}, wave packet isotope separation \cite{R12}, number factorization \cite{R13}, molecular coherent control \cite{R14}, position dependent mass \cite{R15} etc. Underlying mesoscopic quantum superposition holds the secrets for developing quantum technologies and quantum precision measurements through sub-Planck scale structures \cite{Zurek,R16,R17,R18,R21}.

On the other hand, quantum sensing is a highly emerging field for ultracold atoms \cite{Berman,Andrews,Altin,Rasel,Sackett,Rasel2}. To observe FR-phenomena, the systems essentially need to remain stable for sufficiently long time, which is a challenging task in experiments while dealing with mesoscopic quantum objects. Hence, scientists always look for appropriate systems or laboratory environments to achieve the same. Bose-Einstein condensate (BEC) is a highly tunable and coherent wave-packet with large coherence time, making it a promising candidate to observe FR \cite{Greiner,Filip}. Thus, it becomes interesting that FR be precisely reported for Bose-Einstein condensate. Observation of FR-phenomena for matter-wave will considerably intensify the progress of this field. However, there is no exact analysis of FR phenomena in an ultracold atomic system to date.

In this paper, we report the exact structure of the FR phenomena for BEC in a ring-shaped geometry. A ring potential is routinely created by applying a blue-detuned laser beam in the middle of the harmonic trap \cite{Gupta,Ryu1}. BEC loaded in a ring trap has attracted enormous attention in past years \cite{Eckel,Wood,Corman,Beattie,Ryu2,Benakli,Brand,Das,Modugno,Amico}, including persistent flow \cite{Yakimenko1,Bargi,Mateo,White,Moulder}, solitary waves \cite{Mason}, vortices \cite{Piazza,Yakimenko2,Wright} and interference \cite{Berman,Bell,Pedri}. In the following section, we study the condensate dynamics by numerically solving the Gross-Pit\"{a}evskii equation (GPE). The density snapshots at different times reveal the clear signature of FR phenomena in BEC. Illustrations with the autocorrelation function \cite{SG1,Veilande} and quantum carpet \cite{Kazemi,Kaplan} provide a compelling evidence along with the stability analysis. The FR time-scales are evaluated through both theoretical and numerical approaches. Albeit a completely different origin of the current approach, the established exact formula coincides with the standard definition of FR in other quantum systems. The variation of the revival time with the ring-radius, following the theoretical and the computational approaches, are in unsurpassed conformity.

\begin{figure}
\centering
\includegraphics[width=2.7in]{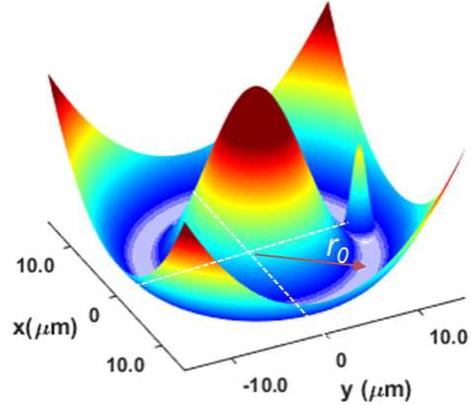}
\caption{Trapping potential is shown which is created by combining a harmonic trap (strength $2\pi\times32.45$Hz) with a Gaussian spike (amplitude
$200\hbar\omega_{\perp}$ and waist $11.35 \mu m$). The initial condensate of width $2.21 \mu m$ is situated at ($0,\;11.85 \mu m$) inside the trap.} \label{potential}
\end{figure}

\section{Solving the Dynamics and Results}

To investigate the dynamics of BEC in a 2D ring-trap, the well-known 3D Gross-Pit\"{a}evskii equation (GPE) \cite{Pethick} is effectively reduced to quasi-2D:
\begin{eqnarray}
\it i \frac{\partial \psi}{\partial t}=-\nabla^2_{x,y}\psi+
 g |\psi|^{2} \psi+V_{ext}(x,y)\psi,\label{2dgp}
\end{eqnarray}
where $\psi$ becomes a function of $x$, $y$, and $t$. $g=4\sqrt{\pi} N a/a_{\perp}$ is the nonlinearity coefficient for the condensate of $N$-atoms of each mass $M$. The $s$-wave scattering length is given by $a$ and the transverse oscillator length is $a_{\perp}=\sqrt{\hbar/(2 M \omega_{\perp}})$ with
$\omega_{\perp}$, being the trap frequency in $z$-direction. The dimensionless form of the above equation is obtained after scaling position, time and energy by $a_{\perp}$, $1/\omega_{\perp}$, and $\hbar \omega_{\perp}$, respectively. The external trapping potential, $V(x,y,t)$, is a combination of a 2D harmonic trap and a Gaussian potential:
\begin{equation}
V_{ext}(x,y)= \frac{1}{4}\omega^{2}r^{2} + V_{0}e^{-2r^{2}/\sigma^{2}}
\end{equation}
with $r^2=x^2+y^2$. To avoid the undesired oscillation in the radial direction, the initial condensate is placed at the exact minimum of the ring ($r_o$), which is obtained by minimizing the resultant potential: $r_{0}= \frac{\sigma}{\sqrt{2}}\sqrt{ln\frac{2V_{0}}{\omega^{2}\sigma^{2}}}$. A Taylor series expansion of the potential around $r_{0}$ provides the effective harmonic trap frequency, $\omega_{e}=2\sqrt{2}\omega r_{0}/\sigma$ \cite{Zhang}.

We solve Eq.~(\ref{2dgp}) by Fourier split step method. Both $x$- and $y$-coordinates are equally divided into $512$ grids with step-size $0.123$. Time grids are having step-size $0.0976$, which makes total $7804$ grids upto the revival time. For illustrating our results, we have considered a BEC of $N$ ($=10^{3}$) $^{85}$Rb atoms with the parameters: $M = 85 a.u.$, $\omega_{\perp}=2\pi\times51$ Hz, $a_{\perp}= 1.077 \mu m$, $a= 58.19\times10^{-11}$m,
$g = 3.829\hbar \omega_{\perp}$ and $\omega = 0.636\times\omega_{\perp}$ \cite{Ryu1}. The trap parameters are $V_{0}=200 \hbar \omega_{\perp}$ and
$\sigma= 11.35 \mu m $, which prepares a ring potential of effective radius $r_{0}=11.857 \mu m $. We have chosen the initial condensate of waist $d_{0}=2.21 \mu m$, placed along the $y$-axis with coordinate $(0,r_{0})$ as shown in Fig.~\ref{potential}. These parameters also enable us to minimize the undesired radial oscillation and breathing inside the ring.

\begin{figure*}
\centering
\includegraphics[width=5.4 in]{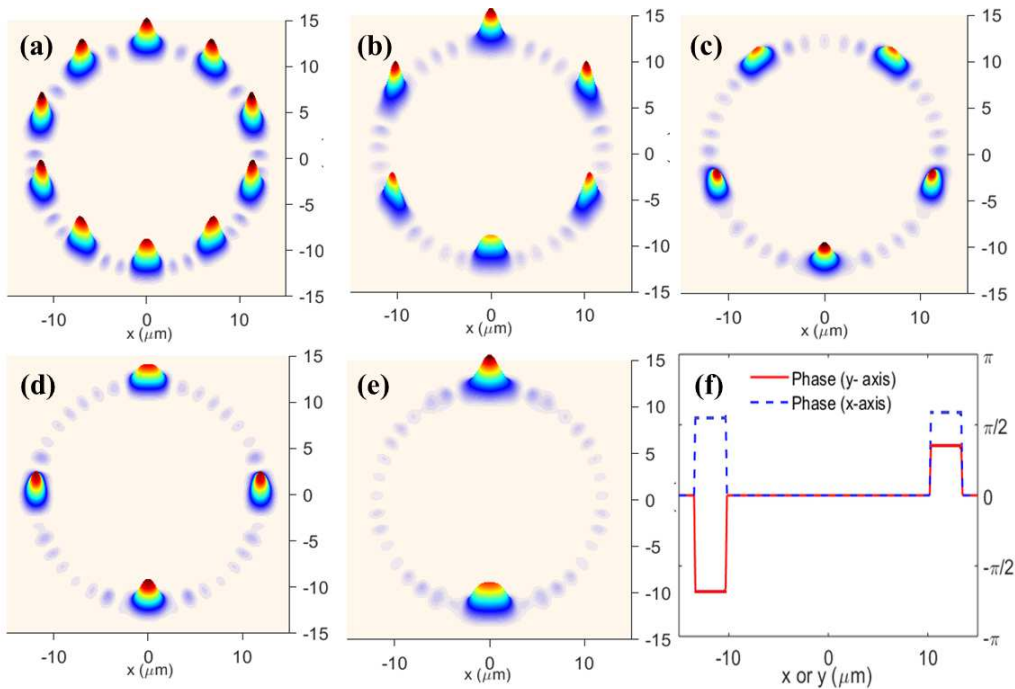}
\caption{ (a-e) Density plots at times $119.80$ms, $198.43$ms, $238.21$ms, $298.11$ms
and $593.89$ms for which the number of petals are $10$, $6$, $5$, $4$ and $2$, respectively.
(f) phase variation at time $298.11$ ms ($s=4$) along $x$-axis (blue dashed line) and $y$-axis (red solid line).} \label{density}
\end{figure*}

\subsection{Condensate density, auto-correlation function and quantum carpet}
As it has been observed in past literature that the initial condensate is expected to disperse along the ring waveguide in both clock-wise and anti-clock-wise directions. Both the counter-propagating components will gradually merge at the opposite pole of the ring $(0,-r_{0})$, producing several closely-spaced spatial interference fringes therein which then gradually disperse. However, spectacular phenomena occur if one observes the condensate for long time. The dispersed clouds reunite again only in some specific time-intervals and give birth to several daughter-condensates which are the replicas of the initial one. All the newly born mini-condensates become similar in shape and size. This phenomena are well-known as fractional revivals. In the due course, all the clouds finally localize again and revive at a particular time, called revival time ($T_{r}= 2.377$s), when it becomes almost identical to the initial condensate. At some fractions of this revival time, several mini-localized condensates are formed. We have recorded the time-fraction and the corresponding number of splits, which reveal the interesting formula,
\begin{eqnarray}
t=\frac{p T_{r}}{q}\label{peatlstime},
\end{eqnarray}
where, $p$ and $q$ are mutually prime integers. It is also observed that the initial condensate splits into $q$ mini-condensates when $q$ is odd and to $q/2$ mini-condensates when $q$ is even. The result is unfolded in Fig.~\ref{density}, where varied time snapshots for condensate density are depicted in Fig.~\ref{density} (a-e), where the initial density splits into (a) $10$-parts at $1/20\; T_{r}$, (b) $6$-parts at $1/12\; T_{r}$, (c) $5$-parts at $1/10\; T_{r}$, (d) $4$-parts at $1/8\; T_{r}$, and (e) $2$-parts at $1/4\; T_{r}$. This is in conformity to the above compelling formula, which clearly reflects the signature of FR-phenomena in this system. For nicely distinguished higher order FR (Fig.~\ref{density} (a)), the radius of the ring has to be sufficient enough and a well-localized initial condensate is also desired. The typical range of ring radius taken in BEC-experiments in literature is already favourable for observing higher order FR.

Figure~\ref{density} (f) depicts the phase-variation along $x$- and $y$-directions for $q=4$ at time $t=298.11ms$ (Fig.~\ref{density} (d)). At this time, four mini-condensates are produced at $x=\pm r_0$, $y=\pm r_0$, when the condensate was initially placed at $y$-axis ($x=0$, $y=r_0$). If we consider the petals on left and right of $x=0$ line, they will always have the same phase because of the symmetry. But, the petals on both sides of $y=0$ line will have different phases. The scenario will be reversed if we initially place the condensate on $x$-axis.

\begin{figure*}
\centering
\includegraphics[width=6.6 in]{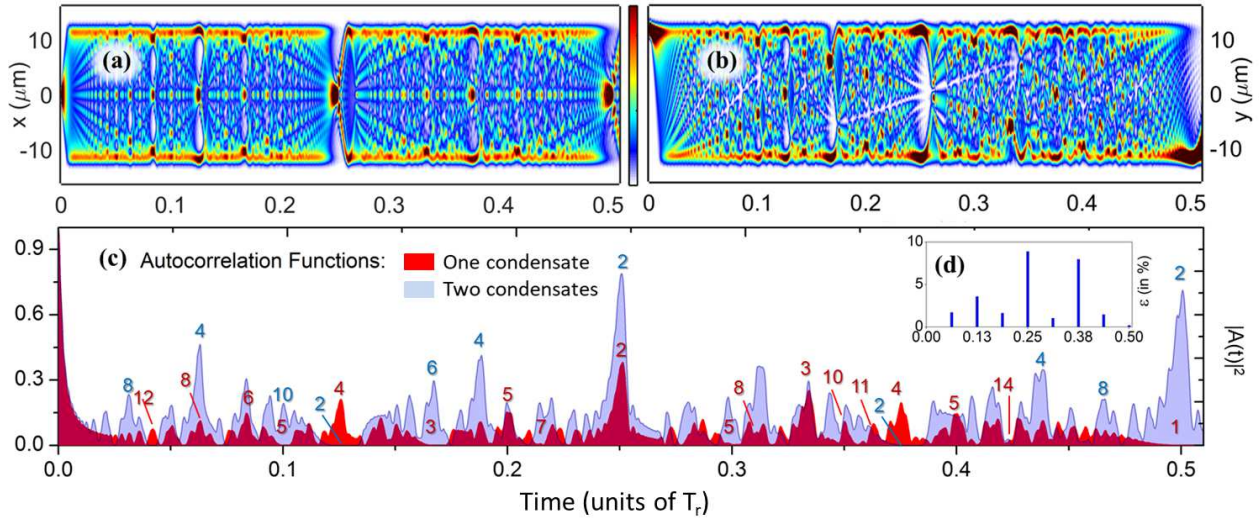}
\caption{Spatio-temporal flow of quantum information or quantum carpets in (a) $x-t$ and (b) $y-t$ planes.
(c) depicts autocorrelation functions for single initial condensate (red line) and two initial condensates (blue line). Numbers indicated against the autocorrelation peaks are the number of splits (denoted by $s$). The stability analysis in presence of noise is shown in (d).}\label{correlation}
\end{figure*}

For a better insight of the FR phenomena, we have investigated the autocorrelation function (AF) and the quantum carpet, which are depicted in Fig.~\ref{correlation}. The autocorrelation function is represented by the modulus square of the overlap between the initial and the final wave functions:
\begin{equation}\label{autocorelation}
|A(t)|^{2}=|\langle\psi_{t=0}|\psi_{t}\rangle|^{2}.
\end{equation}

For pure states, AF is also the time dependent quantum fidelity between the initial and the time-evolved states. In Fig.~\ref{correlation}, we have displayed the variations for $0\leq t\leq T_{r}/2$, beyond which ($T_{r}/2\leq t\leq T_{r}$) it is the mirror image of the first half. The well-defined peaks of the AF signify FR instances. These instances are designated in Fig.~\ref{correlation}(c) with labelling of the number of mini-condensate.
Dark-shaded (red) plot in Fig.~\ref{correlation}(c) is the AF of the case under study. Fig.~\ref{correlation}(a) and ~\ref{correlation}(b) are displaying the quantum carpets along the $x$- and $y$-quadratures, respectively. Quantum carpet is a popular nomenclature for the space-time flow of the quantum information. Localizations of the condensate densities are visible in both the quadratures at different FR times. When evaluating the quantum carpet along $x$-direction, we integrate over $y$ and vice-versa. Additionally, the space-time rays originating from $t=0$ and $T_{r}/2$ signify different velocities of the different constituent waves (the cause of dispersion). It is worth notifying that at time $t=0.25\;T_r$, the initial space-time spots revive at $x=0$, but splits into two parts along $y$ ($=\pm r_0$) and the AF gets a large peak. However, at time $t=0.125\;T_r$, the condensate splits into four parts and becomes symmetric in $x$- and $y$-directions (Fig.~\ref{density} (d)), leaving behind the same structure (three bright spots) in both the quantum carpets (Fig.~\ref{correlation}(a) and (b)). AF at this time also produces a peak. One can explain all the other FR instances and correlate Fig.~\ref{correlation} nicely with Fig.~\ref{density}. To indicate the difference, we have also included a lighter shaded (blue) plot in Fig.~\ref{correlation}(c) for the case where one starts with two condensates, symmetrically situated in the opposite poles of the toroid. In this case, the structures are more prominent with the AF peaks of higher magnitude and the revival time-scale becomes half of that for the first case.

\subsection{Theoretical Derivation of the time-scale formula}

To investigate the formation mechanism of the FR-phenomena, we begin by taking the expression of two counter propagating Gaussian wave-packets separated by a distance $D$ \cite{Pethick}:
\begin{eqnarray}
\Phi_{\pm}(d,t)&=&N_{\pm} exp\Big[- \frac{(d\pm D/2)^{2}(1+ i 2t/\omega_{0}^{2})}{2\omega_{t}^{2}}\Big]\label{wavepacket},
\end{eqnarray}
where $N_{\pm}$ is the normalization constant, $\omega_{0}$ is the initial width of the wave-packet and $\omega_{t}=\sqrt{\omega_{0}^{2} + (2t/\omega_{0})^{2}} $ is the width after time $t$. Such spreading of the wave-packet becomes eventually linear with time ($\omega_t=2t/\omega_0$) for long time evolution.

Let us imagine the present case, where the initial condensate is placed at $(0,r_{0})$ (Fig.~\ref{potential}). It can be modeled by two wave-packets separated by $D=2\pi r_0$, where $r_0$ is the effective radius of the ring. The two clouds are coexistent initially and subsequently spreading clock-wise and counter-clockwise along the minimum of the ring. After propagating through the ring, the two identical counter-propagating clouds start interfering at $(0,-r_{0})$. The interference maxima are obtained from the oscillating term, $\cos(2Ddt/\omega_{0}^{2}\omega_{t}^{2})$. Hence, the fringe separation is given by
\begin{equation}
\Delta d=\pi\omega_{0}^{2}\omega_{t}^{2}/(D t),\;and \; \Delta d^{\prime}=4\pi t/D,
\end{equation}
where the later is the effective fringe separation for the long time evolution \cite{Andrews,Pethick}. Imagine the whole string as a ring of radius $r_0$ and the initial wave-packets splits into $s$ number of mini replicas in long time evolution. Due to the circular symmetry of the waveguide, the separation between the two maxima in the course of long time evolution is evaluated from the relation, $\Delta d^{\prime}\times s=2\pi r_0 m$, where $m$ is a nonzero integer, signifying the number of times the condensate circles the origin in a given time. Making use of the expressions of $D$ and $\Delta d^{\prime}$, we obtain
\begin{equation}
t=\pi r_0^2 \frac{m}{s}. \label{FR}
\end{equation}

\begin{figure}
\centering
\includegraphics[width=2.8 in]{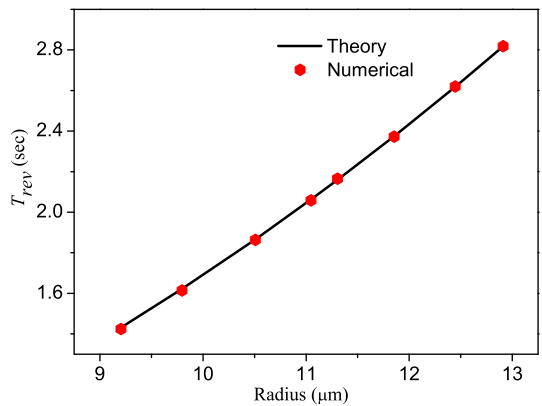}
\caption{Comparison of the theoretical values (from Eq.~(\ref{FR})) and numerical values of the time scales with the size of the ring-trap. They are in very good agreement.}\label{trevradius}
\end{figure}

Condensate will revive ($s=1$) at times, $\pi r_0^2$, $2\pi r_0^2$, $3\pi r_0^2$ and so on. However, for the odd multiples of $\pi r_0^2$, the wave-packet revives in shape, but not in position as it revives at the opposite end of the ring. If one defines the revival time when the wave-packet revives both in shape and position, then it becomes $T_{rev}=2\pi r_0^2$. In addition, obtaining various number of mini-wave-packets will require $m$ and $s$ to be mutually prime integers. Here, $m$ and $s$ grasp identical meaning of $p$ and $q$ in Eq.~(\ref{peatlstime}) and amazingly, this formula is identical to what has been observed in other nonlinear quantum systems. We also compare the revival time calculated from the theoretical formula (Eq.~(\ref{FR})) with the revival time obtained from numerical result. The quadratic dependence of the revival time with the ring radius is confirmed in Fig~\ref{trevradius}. For the experimental parameters considered in this work, $r_0$ is $11.857\mu m$, for which the revival time obtained as $T_r = 2.377 s$. This coordinate is also precisely lying on the curve (Fig~\ref{trevradius}). It is worth mentioning that for each point on this curve, a proper choice of the trap and initial condensate parameters are required to obtain a perfect FR. Additionally, a slight change in the ring radius by $0.01\mu m$ is associated with the change in revival time $4 ms$.

\subsection{Stability analysis}
Investigating the stability of the condensate with time is inevitable, particularly when one studies the long time evolution. The stability analysis has been carried out by the conditionally stable, split-step fast Fourier transformation (SSFFT) method, which is an established numerical method and widely used in numerical simulation of non-linear wave propagation, mainly in the context of nonlinear Schrödinger equation (NLSE) \cite{Weideman,Szc}.
Here, we present the stability analysis for the previously mentioned parameters in this manuscript and mix white noise ($\Sigma (x,y)$) to the initial wavefunction with mean zero and amplitude $5\%$ of the maximum value of the condensate density at a particular time.
\begin{equation}
\psi_{t=0}=\psi_{t=0}+ \Sigma (x,y),
\end{equation}

The condensate wave function with noise is evaluated after a certain time and corresponding autocorrelation function (AF) is calculated from Eq.~(\ref{autocorelation}). Then a percentage deviation of the AF (with noise) from the AF (without noise) is calculated which is elucidated in Fig.~\ref{correlation}(d). The times for which these steps are repeated are $1/16$, $1/8$, $3/16$, $1/4$, $5/16$, $6/16$, $7/16$ and $1/2$ (times are scaled by $T_r$), which stands for an adequate stability of the condensate in the course of its entire time evolution. We have repeated the same study for other ranges of parameters and found the solution sufficiently stable.

\section{Conclusion}
In conclusion, we have reported FR phenomena for a 2D system of BEC in a ring-shaped waveguide. Condensate splitting in the 2D-real space is depicted through density snapshots, autocorrelation function and spatio-temporal structures in quantum carpets. We have eliminated unwanted breathing of the cloud in the radial direction by appropriately choosing the initial condensate. Proper tuning of the trap parameters (harmonic and Gaussian) within the experimental parameter regime is important to prepare the minima of the ring in harmonic shape. We have also shown the phase variation in Fig.~\ref{density}(f) at $t=T_{r}/8$, which imparts an idea of real space interference patterns in a time-of-flight measurement, which is being studied in a separate work. We identify the time scales quantitatively and compare the same with our theoretical model where we obtain an exact formula. A numerical stability analysis makes the results more viable for experimental realization. The variation of the revival time with the radius of the ring fosters quantum sensing applications. The work can be extended for 2D-disordered lattice for which the strong localization of condensate can be utilized to get better visibility during FR \cite{Roy-LPL,Roy-SR}. This work will also be a suitable guideline for a variety of other applications towards quantum information, quantum metrology, quantum logic gates and decoherence.

\end{document}